\documentclass[journal]{IEEEtran}
\usepackage{graphicx}
\usepackage{comment}
\usepackage{amsmath,amssymb} % define this before the line numbering.
\usepackage{color}
\usepackage{epsfig}
\usepackage{graphicx}
\usepackage{booktabs}
\usepackage{enumitem}
\usepackage{subfigure}
\usepackage[pagebackref=true,breaklinks=true,colorlinks,bookmarks=false]{hyperref}
\usepackage{multirow}

\ifCLASSINFOpdf
\else
\fi
\begin{document}
\title{Multi-Texture GAN: Exploring the Multi-Scale Texture Translation for Brain MR Images}

%%%------------------
\author{Xiaobin~Hu,
        Yu~Zhao,
        Anjany~Sekuboyina,
        Bjoern~Menze% <-this % stops a space

% \author{Anonymous
%         }

%%%-------------------     
% \author{Xiaobin~Hu,~\IEEEmembership{Member,~IEEE,}
%         Yu~Zhao,
%         Anjany~Sekuboyina,
%         Bjoern~Menze,~\IEEEmembership{Life~Fellow,~IEEE}

% \thanks{M. Shell was with the Department
% of Electrical and Computer Engineering, Georgia Institute of Technology, Atlanta,
% GA, 30332 USA e-mail: (see http://www.michaelshell.org/contact.html).}% <-this % stops a space
% \thanks{J. Doe and J. Doe are with Anonymous University.}% <-this % stops a space
% \thanks{Manuscript received April 19, 2005; revised August 26, 2015.}
% \IEEEcompsocthanksitem X. Hu  and M. Bjoern are with the Department
% of Informatics, Technische Universit\"at M\"unchen, Germany, 80333 Munich, Germany (email:xiaobin.hu,bjoern.menze@tum.de)

%%----------------
\thanks{X. Hu, Y. Zhao, A. Sekuboyina and M. Bjoern are with the Department
of Informatics, Technische Universit\"at M\"unchen, Germany, 80333 Munich, Germany (email:xiaobin.hu@tum.de, yuzhao90@outlook.com, anjany.sekuboyina@tum.de, bjoern.menze@tum.de)}}
% 

%%---------------


% \markboth{IEEE Journal of Biomedical And Health Informatics}%
\markboth{submitted to journal }%
{Shell \MakeLowercase{\textit{et al.}}: Bare Demo of IEEEtran.cls for IEEE Journals}



\maketitle

\begin{abstract}
Inter-scanner and inter-protocol discrepancy in MRI datasets are known to lead to significant quantification variability. Hence image-to-image or scanner-to-scanner translation is a crucial frontier in the area of medical image analysis with a lot of potential applications. 
Nonetheless, a significant percentage of existing algorithms cannot explicitly exploit and preserve texture details from target scanners and offers individual solutions towards specialized task-specific architectures. In this paper, we design a multi-scale texture transfer to enrich the reconstruction images with more details. Specifically, after calculating textural similarity, the multi-scale texture can adaptively transfer the texture information from target images or reference images to restored images. Different from the pixel-wise matching space as done by previous algorithms, we match texture features in a multi-scale scheme implemented in the neural space. The matching mechanism can exploit multi-scale neural transfer that encourages the model to grasp more semantic-related and lesion-related priors from the target or reference images. We evaluate our multi-scale texture GAN on three different tasks without any task-specific modifications: cross-protocol super-resolution of diffusion MRI, T1-Flair, and Flair-T2 modality translation. Our multi-texture GAN rehabilitates more high-resolution structures (i.e., edges and anatomy), texture (i.e., contrast and pixel intensities), and lesion information (i.e., tumor). The extensively quantitative and qualitative experiments demonstrate that our method achieves superior results in inter-protocol or inter-scanner translation over state-of-the-art methods.
\end{abstract}


% \fi
\IEEEpeerreviewmaketitle

\section{Introduction}
Currently, with the use of high-field scanners, Magnetic Resonance Imaging (MRI) has become prominent in the domain of neuro-radiology, since the MRI contrasting agent is less likely to induce an allergic behavior that may occur when iodine-based agents are used for x-rays and CT scans. Furthermore, in contrast to other imaging techniques, MRI offers clear and accurate images of soft-tissue structures. Consequently, they are widely used for the purposed of the clinical treatment for diagnosis and therapy. The MRI modalities capture various characteristics of the underlying anatomy and represent the different texture and contrast knowledge. For clinical diagnosis, three major modalities of MR imaging are widely referenced: Flair (Fluid attenuation inversion recovery), T1 (Spin-lattice relaxation),and T2 (Spin-spin relaxation). Although obtaining all modalities is beneficial for MRI analysis and diagnosis, it is still arduous to collect fully complete multi-modality MR images due to the failure during the process of scanning (e.g., motion artifacts) and the lack of the equipment. Additionally, higher-field MRI scanners facilitate clinic diagnosis and prognosis by producing image with detailed anatomical information and higher signal-to-noise ratio. But the cost-prohibitive aspect of high-field scanners makes them less accessible. These obstacles motivate the proposal of cross-modality and cross-scanner image synthesis techniques to resolve them via efficient data infilling and re-synthesis.

 %%--------------
\begin{figure*}
\centering
\includegraphics[width=0.9\textwidth]{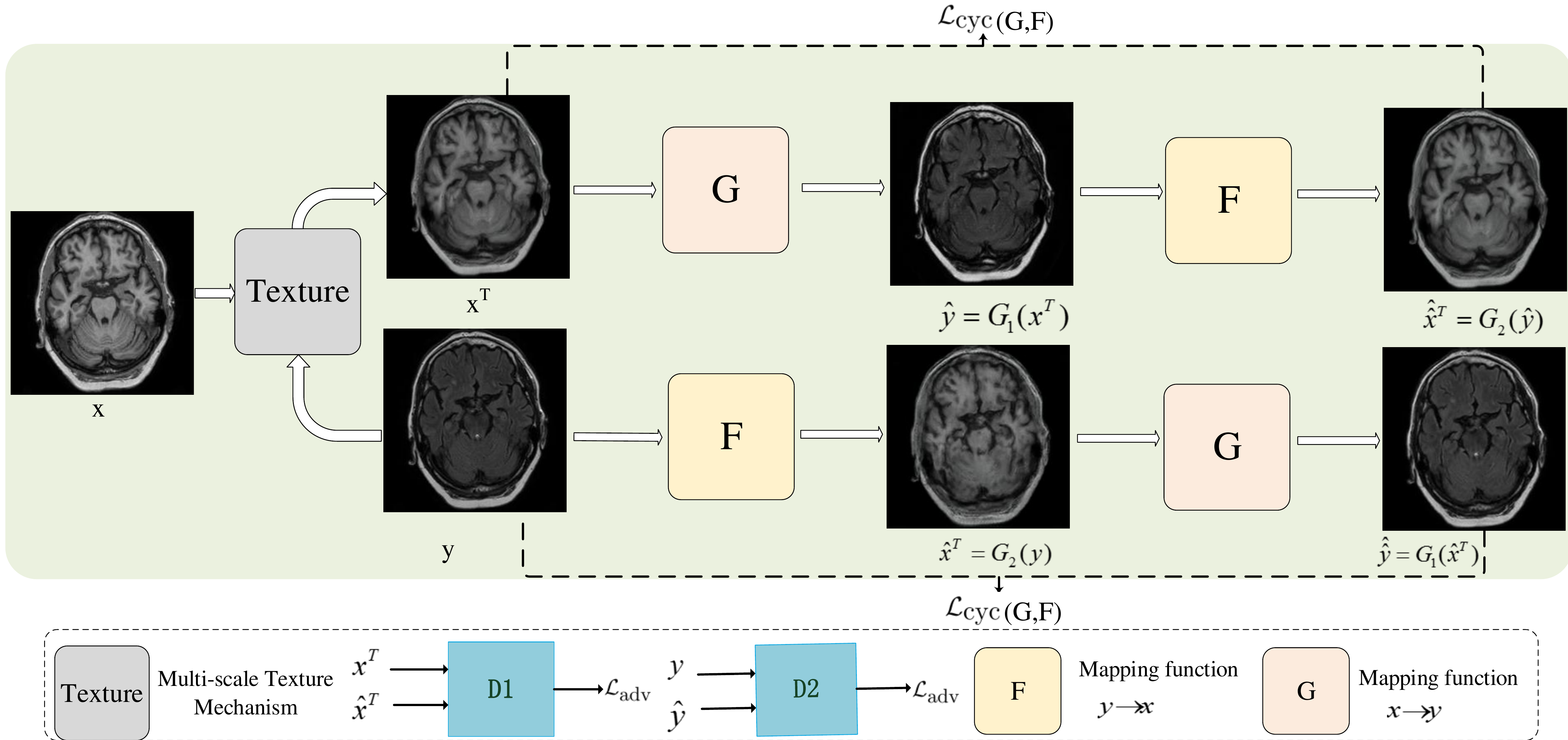}
\caption{Architecture of the proposed multi-scale texture GAN model. Our model contains two parts: 1) The multi-scale texture mechanism to exploit the similarity and generate the high-level texture details. 2) The cycle-consistent training fashion to reconstruct super-resolution MRI by using the estimated texture priors. }\label{fig:fig1}
\end{figure*}   
%%------

%  computer vision and machine learning techniques have been verified effective for a variety of medical image tasks. 

Recently, the development of deep learning methods, especially the convolutional neural network (CNN) promoted the dramatic improvement in a variety of medical applications, such as lesion classification and detection \cite{classify_1,classify_2}, segmentation \cite{seg_1,seg_2} and image resolution enhancement \cite{sr_1,sr_2}. This also boosted the development of some approaches for medical image synthesis and enhancement \cite{xiang2018deep,bahrami2020new}. However, a huge milestone of the image translation is when the generative adversarial networks (GANs) \cite{gan} was proposed. GANs are not only widely applied into the images segmentation \cite{gan8} and classification \cite{gan7} but also perform a breakthrough role in the image-to-image translation. GANs are the state-of-the-art algorithm in the aspect of recovering unprecedented levels of image realism \cite{gan1}. GANs and its extensions \cite{gan10,gan3} were to learn the distribution of synthesis data from the source data, with a large portion of work focusing on the theoretical and architectural analysis \cite{gan4,gan5}. Ledig et al. \cite{gan6} presented a generative adversarial network for image super-resolution, also named SRGAN, by designing a perceptual loss function that combined with a content loss and a typical adversarial loss. Isola et al. \cite{gan9} proposed a pix2pix network by investigating conditional adversarial networks for a general image-to-image translation task. The pix2pix networks tailored a novel loss to train a mapping from source images to target images without the modification of different loss formulation for different specific translation tasks. Unpaired image translation algorithm was presented by Zhu et al. \cite{gan10} by designing a dual cycle consistency loss from source domain to target domain. Furthermore, GANs have been effectively introduced into the medical image field especially for image-to-image tasks. For the purpose of not generating blurry images, Nie et al. \cite{nie2018medical} proposed an image-gradient-difference-based loss function using a deep convolutional adversarial network to synthesis 7T MRI from 3T MRI images. Armanious et al. \cite{armanious2019unsupervised} presented a Cycle-MedGAN for PET-CT translation and the correction of motion artifacts utilizing novel non-adversarial cycle losses to guide the framework to optimize the texture and perceptual discrepancies in the translated images. A new generator of GAN-based architecture CasNet was presented in \cite{armanious2020medgan} for enhancing the sharpness of the output images using the progressive refinement by encoder-decoder structure. The above approaches are an overview of different applications of adversarial networks on natural and medical translation tasks. However, the above GANs-based translation networks ignore the contrast and texture transformation between source and target domains.

Neural texture transfer is a useful mechanism to adaptively transfer texture from the reference images to the source images \cite{zhang2019image}. The neural algorithm that separates and then recombines the content and style of images was presented by Gatys et.al \cite{gatys2016image}. Essentially, it is a texture transfer method that contains the process of texture synthesis by feature representations and local texture matching. But the process of matching and swapping texture features is only based on single-scale aspect and perform weakly on texture transfer and fail on scale-up problem (e.g., super-resolution) caused by different spatial size. Inspired by this idea \cite{johnson2016perceptual,gatys2016image}, we propose a multi-scale texture mechanism to synthesize the texture representations of the source images and exploit the GANs-based algorithm to better amend the content features of the source images guided by the reference images. We conclude and emphasize the main contributions as following: 
\begin{itemize}
% \vspace{-0.3cm}
    \item [1)] We explore a general texture-based GANs method for MR image synthesis, breaking the performance barrier (e.g., lack of the texture priors) and the requirement barrier of identical spatial size in medical image enhancement and translation. Consequently, our multi-scale texture GANs extend its applications into the super-resolution task with non-identical spatial size between the inputs and outputs.
    \item [2)] We propose a multi-scale mechanism for the texture transfer between source and target domain. Compared with previous texture transfer algorithms, this mechanism adaptively transfers the texture from the multi-scale matching according to their texture similarity.
    \item [3)] We demonstrate the performance in three crucial tasks: i) cross-protocol super resolution of diffusion MRI, ii) T1-Flair MRI cross-modality translation, iii) Flair-T2 MRI cross-modality translation. The proposed network achieves better performance compared to state-of-the-art algorithms. 
\end{itemize}

Firstly, we present in detail each component of the multi-scale texture GANs-based algorithm. We then describe the specific datasets and evaluate the performance of our method in the cross-protocol and cross-modality tasks. Finally, we compare with very recent network frameworks (U-net \cite{ronneberger2015u}, CycleGAN \cite{gan10}, Texture-CycleGAN) and show the superiority of the proposed network (multi-scale texture CycleGAN).

% performs much better than the state-of-the-art image transfer methods in terms of structural similarity (SSIM), peak signal-to-noise ratio (PSNR) and mean-square error (MSE).

%%%-----
\begin{figure*}[t]
\centering
\includegraphics[width=1\textwidth]{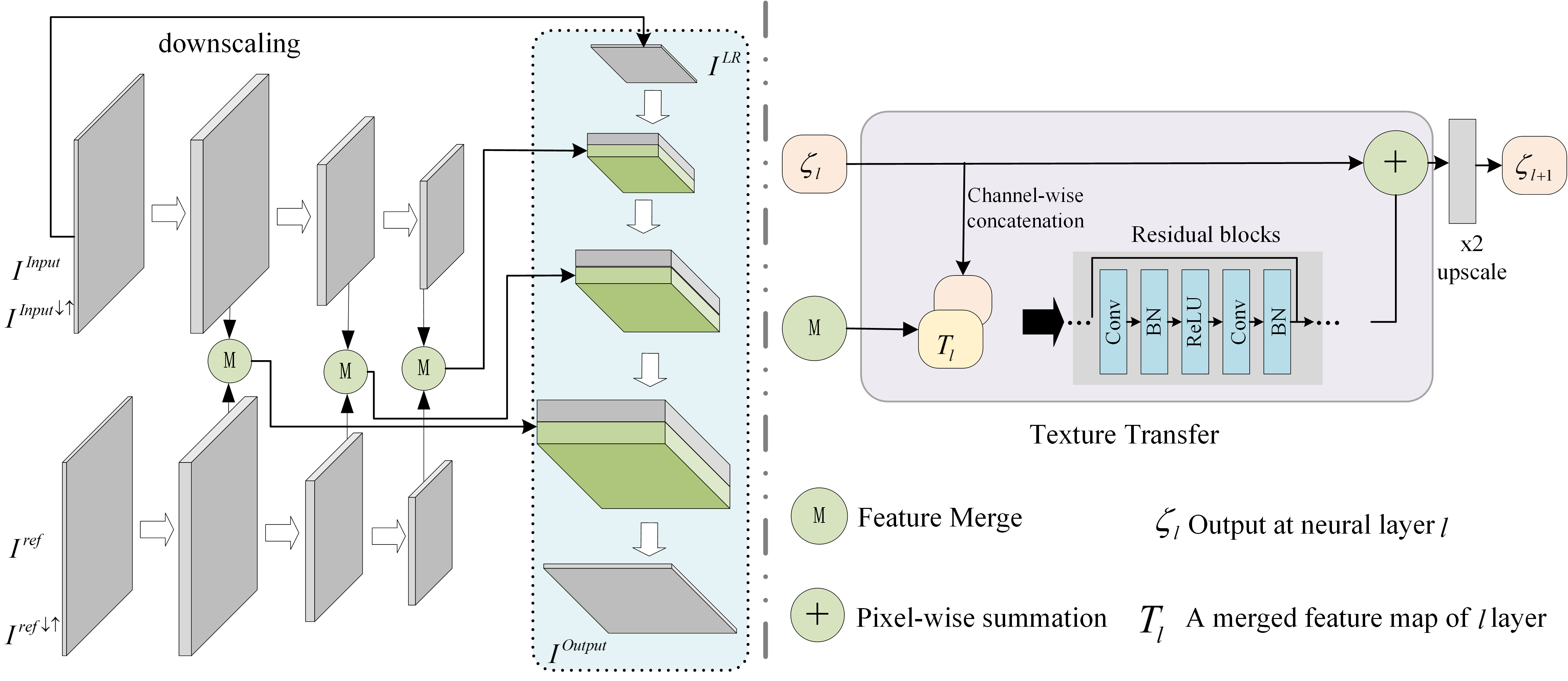}
\caption{The details of multi-scale texture mechanism with feature merging and texture transfer. The $I^{Ref\uparrow\downarrow}$, a blurry image down-sampling and up-sampling the high-resolution image, is used for matching while the high-resolution  $I^{Ref}$ is used in merging to ensure the HR information from the original reference is preserved. }\label{fig:fig2} 
\end{figure*}
%%%------

\section{Method}
The proposed framework aims to synthesize the target image (ground truth images or scanners) from its source images (input images or scanner) and the reference images (similar texture knowledge to the target images) using the multi-scale texture transfer mechanism and cycle-consistency training fashion, shown in Figure \ref{fig:fig1}. The multi-scale texture mechanism in Figure \ref{fig:fig2} transfers related texture and suppresses the irrelevant textures by considering and calculating high-level (semantic) and low-level (textural and contrast) similarity between the input images and reference images. In addition, we regularize on the texture consistency between the output and the reference image to minimize the pixel and perceptual distance.

\subsection{Feature Matching and Swapping}
The first step of multi-scale texture transfer mechanism is the feature search and swapping over the whole reference images to capture the locally similar texture to replace the features of original texture from input images to enhance the texture knowledge of final outputs. To build up multi-scale texture transfer which gradually upscales the inputs, we firstly use down-sampling on input images to obtain down-scaled low-resolution images as input. In order to dismiss the constraint on the global structure of the reference images, we match and compare the local patches between input images and down-up reference images (blurry ref images) which got by sequentially down-scaled and then up-scaled with the same factor. Since input and reference images are different in texture, color, and contrast, we match them and calculate their similarity in the neural feature space based on the pre-trained VGG network to focus on the textural and structural information. The inner product is used to measure the similarity of neural space:
\begin{equation}
S_{i,j}=\left<P_{i}(\phi(I^{Input})), \frac{P_{j}(\phi(I^{Ref\uparrow\downarrow}))}{	\left \| P_{j}(\phi(I^{Ref\uparrow\downarrow}))\right\|} \right>
\end{equation}
Where $S_{i,j}$ means the similarity of the $i$-th patch of input and the $j$-th patch of reference  and $\phi$ are texture features in neural space. $P_{i}$ and $P_{j}$ are sampling the $i$th patch and $j$th from neural feature map. To select the best match over whole $j$, the ref patch feature is normalized as following: 
\begin{equation}
S_{j}=\phi(I^{Input})\ast \frac{P_{j}(\phi(I^{Ref\uparrow\downarrow}))}{	\left \| P_{j}(\phi(I^{Ref\uparrow\downarrow}))\right\|}
\end{equation}
Where $S_{j}$ means the similarity map of the $j$-th patch of reference , and $\ast$ represents the correlation operation. After calculating the similarity score, a swapped feature map $T$ is built to stand for the texture-enhanced input image. Each patch of enhanced input features $T$ centered at coordinate $(x,y)$ is expressed as:
\begin{equation}
P_{w(x,y)}(T)=P_{\mathop {\arg\max S_j(x,y) }\limits_{j\in Ref}}\phi(I^{Ref})
\end{equation}
Where $\mathop {\arg\max S_j(x,y) }\limits_{j\in Ref}$ aims to select the texture patch of the reference image corresponding to the maximum similarity score. Note that high resolution raw Ref $I^{Ref}$ is used in swapping while the blurry images are just for matching. 

\subsection{Multi-scale texture transfer mechanism}
The multi-scale texture transfer mechanism is tailored by combining the multi-scale swapped texture feature maps with different feature layers of a deep generative network.
% The multi-scale texture transfer model is designed by merging the multi-scale swapped texture feature maps into a base deep generative network at different feature layers corresponding to various scale.
The generative network is constructed by the multi-scale residual block. Each scale block is incorporated with the corresponding swapped feature map. The output $\xi_{\ell}$ at layer $\ell$ is expressed in a recursive way: 
% recursively
\begin{equation}
\xi_{\ell}=[\rm{Res}(\xi_{\ell-1}\oplus T_{\ell-1})+\xi_{\ell-1}]\uparrow_{2\times}
\end{equation}
% Where $I^{Output}$ is the output of the network and $\ell$ denotes the current scale. 
where $\oplus$ is the channel-wise concatenation, and $\rm{Res(\cdot)}$ is the residual block. After recursive process, the output of network is obtained and defined as $I^{Output}$.
We consider the texture difference between  $I^{Output}$ and $I^{Ref}$
by defining a texture loss $L_{tex}$ as:
\begin{equation}
\begin{split}
\mathcal{L}_{tex}=\sum_{\ell}\lambda_{\ell}	\left \| 
Gr(\phi_{\ell}(I^{Output})\cdot S^{\ast}_{\ell})-Gr(T_{\ell}\cdot S^{\ast}_{\ell})
\right\|_{F},
\end{split}
\end{equation}
where $Gr(\cdot)$ denotes the Gram matrix. $\lambda_{\ell}$ is a normalization parameter corresponding to the feature size of scale $\ell$. $S^{\ast}_{\ell}$ is a similarity map of the best matching score for all LR patches. 

\subsection{Cycle-consistent adversarial training}
The proposed network can capture the texture priors after multi-scale transfer mechanism. In order to better incorporate the priors, we utilize the cycle-consistent adversarial training fashion \cite{gan10} to boost the performance of image translation. Following \cite{gan10}, our goal is to learn the mapping between X (images $ \{ x_i \} \in X$) and Y (images $\{y_i\}\in Y$) domains. Two mapping functions are expressed as $G$: $X \rightarrow Y$ and $F$: $Y \rightarrow X$. We employ adversarial discriminators $D_X$ to distinguish between images $\{x\}$ and corresponding translated images $\{F_{y}\}$ and $D_Y$ to discriminate between images $\{y\}$ and corresponding translated images $\{G(x)\}$, respectively. For the mapping $G$: $X \rightarrow Y$ and its discriminator $D_Y$, the objective $\mathcal{L}_{adv}(G,D_Y)$ is defined as:
\begin{equation}
\begin{split}
\mathop{\min}\limits_{G} & \mathop{\max}\limits_{D_Y} \mathcal{L}_{adv}(G,D_Y) \\
&= \mathbb{E}_y[\log D_Y(y)] +\mathbb{E}_x[\log (1-D_Y(G(x)))]
\end{split}
\end{equation}

% \begin{equation}
% \begin{split}
% \mathop {\min }\limits_{\theta _G^{}}&{\mathop {\max }\limits_{\theta _D^{}} ^{}}\xi \left( {\theta _G^{},\theta _D^{}} \right)  \\
% &=\frac{1}{N}\sum_{n=1}^{N}\ell_{mae}{f_{D}(x_{n}\bullet G(x_{n}))},f_{D}(x_{n}\bullet y_{n}))
% \end{split}
% \end{equation}

where $G$ aims to construct fake images $\{G(x)\}$ close to domain $Y$ while $D(Y)$ tries to distinguish between the translated samples $\{G(x)\}$ and real samples $\{y\}$. The procedure is concluded as a min-max optimization task over the adversarial loss function. For the mapping ($F$: $Y \rightarrow X$ and its discriminator $D_X$), the analogous loss is explained as:

\begin{equation}
\begin{split}
\min \limits_{F} & \max\limits_{D_X} \mathcal{L}_{adv}(F,D_X) \\ = &\mathbb{E}_x[\log D_X(x)]+\mathbb{E}_y[\log (1-D_X(F(y)))]
\end{split}
\end{equation}

Due to a lack of sufficient constraints, the model may suffer the mode collapse. Consequently, except for  the adversarial loss, we add the pixel-wise cycle-consistency loss into generators to ensure two generator networks should invert each other in a cycle fashion (i.e., $x \rightarrow G(x) \rightarrow F(G(x)) \approx x$ and $y \rightarrow F(y) \rightarrow G(F(y)) \approx y$). The behavior is incentivized by a cycle consistency loss:
\begin{equation}
\begin{split}
\mathcal{L}_{cyc}& (G,F) \\ = & \mathbb{E}_x[\Arrowvert (F(G(x))-x) \Arrowvert _1]+\mathbb{E}_y[ \Arrowvert (G(F(y))-y) \Arrowvert _1]
\end{split}
\end{equation}

For the cycle-consistent framework, a combination of the adversarial and cycle-consistency losses is employed as a final min-max optimization objective:
\begin{equation}
\begin{split}
\min \limits_{G,F} & \max \limits_{D_X,D_Y}  \mathcal{L}(F,D_X) \\ =& \mathcal{L}_{adv}(G,D_Y)+\mathcal{L}_{adv}(F,D_X)+ \lambda_{cyc}\mathcal{L}_{cyc}(G,F) 
\end{split}
\end{equation}

where $\lambda_{cyc}$ is the weight given for the cycle-consistency part.

\section{Materials}
% This section illustrates the datasets and evaluation metrics to evaluate the quality of image translation.

\subsection{Datasets}
We evaluate the performance of the multi-scale texture GAN network on three tasks and each task is implemented using corresponding brain MRI dataset. For the first task of cross-protocol super-resolution problem, diffusion MRI data (MUSHAC) \cite{tax2019cross} is used. The HR images are obtained by state-of-the-art diffusion MRI acquisition by Connectom scanner with voxel size (1.2$\times$1.2$\times$1.2 $mm^3$), and the corresponding LR images are scanned by the standard acquisition of Prisma scanner with a larger voxel size (2.4$\times$2.4$\times$2.4 $mm^3$). Nine subjects are used as training and one subject for testing. For the task of T1-Flair cross-modality translation, the public WMH dataset \cite{kuijf2019standardized} is employed and randomly divided into training (sixteen patients) and test patients (four patients). For the last task of Flair-T2 cross-modality translation,  twenty-eight patients in BRATS \cite{menze2014multimodal,bakas2017advancing,bakas2018identifying} are randomly selected as the training while seven patients are as test data. 

\begin{figure*}[htbp]\scriptsize
	\begin{center}
		\tabcolsep 1pt
		\begin{tabular}{@{}cccccc@{}}
			\includegraphics[width=0.33\textwidth]{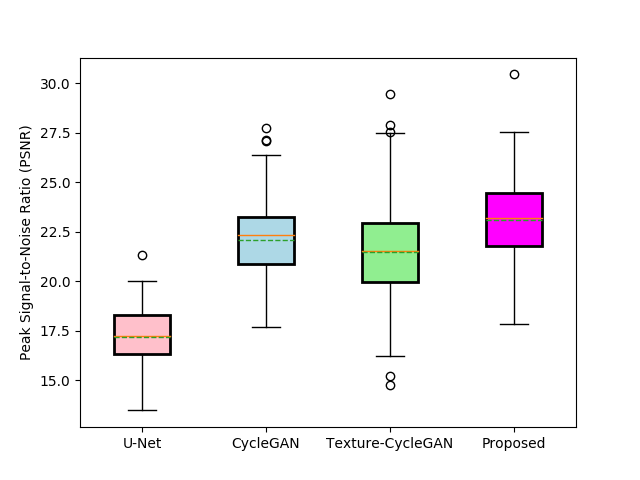}  & 
			\includegraphics[width=0.33\textwidth]{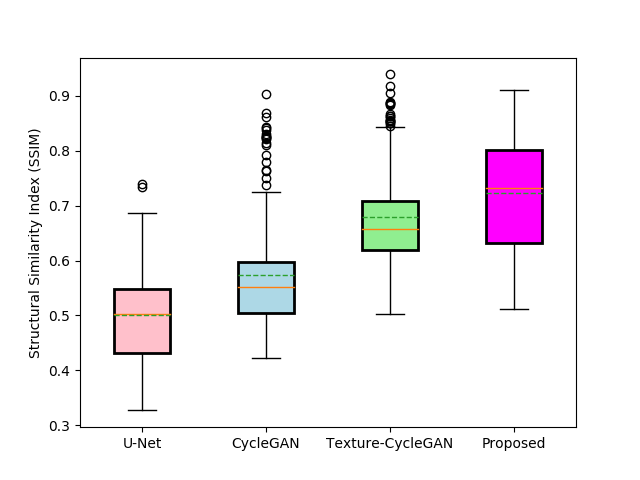} & 
			 \includegraphics[width=0.33\textwidth]{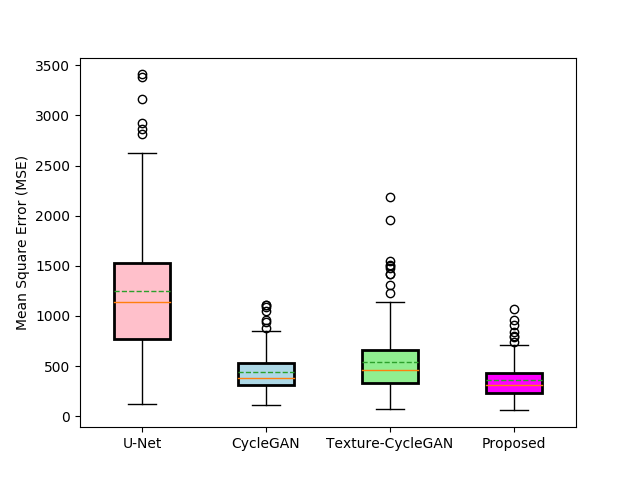}  \vspace{-2mm}  \\
			 (a)   & (b)  & (c)  \\
	
			\includegraphics[width=0.33\textwidth]{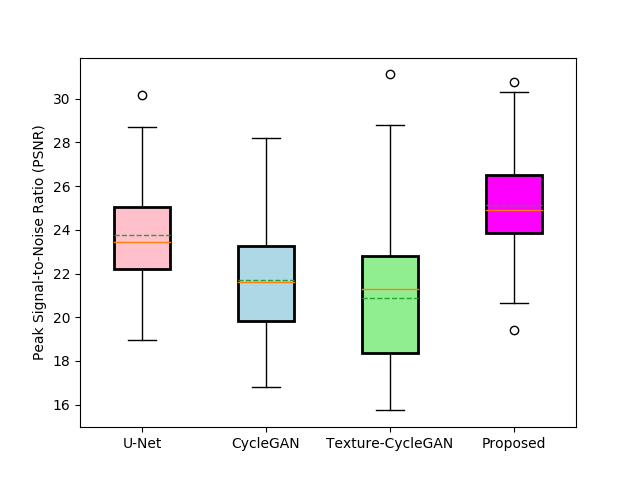}   &
 
			\includegraphics[width=0.33\textwidth]{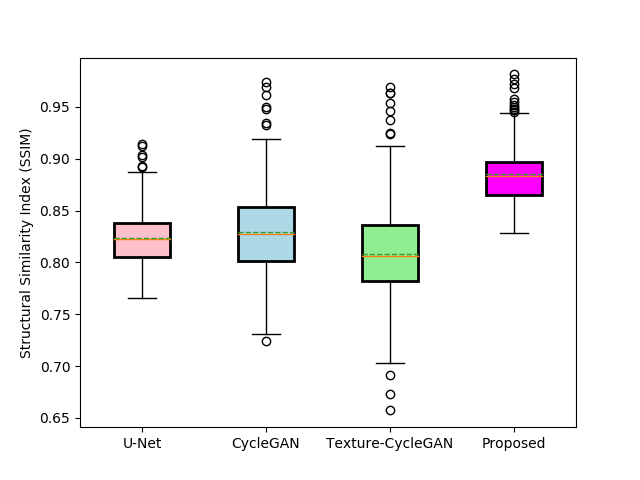}  & 
			\includegraphics[width=0.33\textwidth]{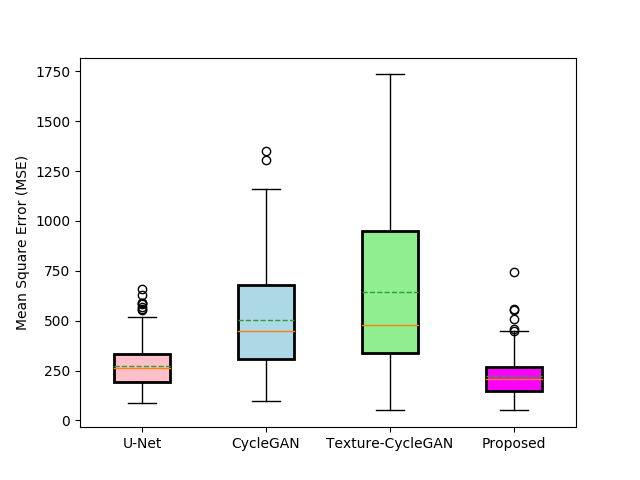} \vspace{-2mm} \\
		(d) & 	(e)  & (f)  \\
		
		\includegraphics[width=0.33\textwidth]{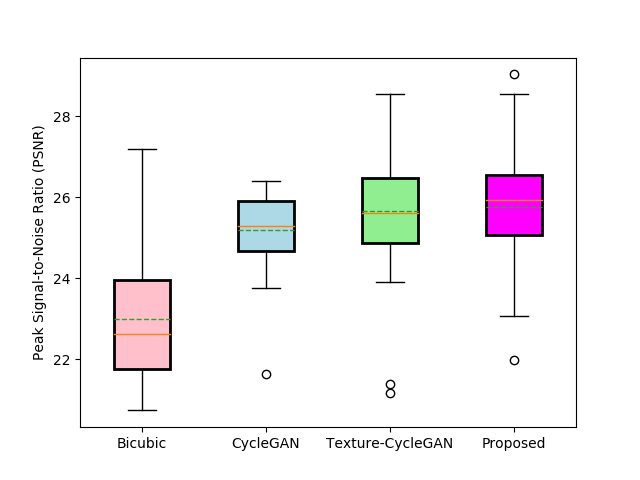}   &
 
		\includegraphics[width=0.33\textwidth]{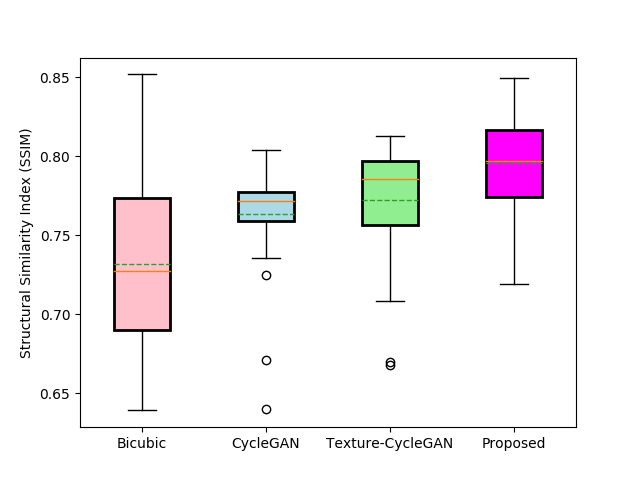}  & 
		\includegraphics[width=0.33\textwidth]{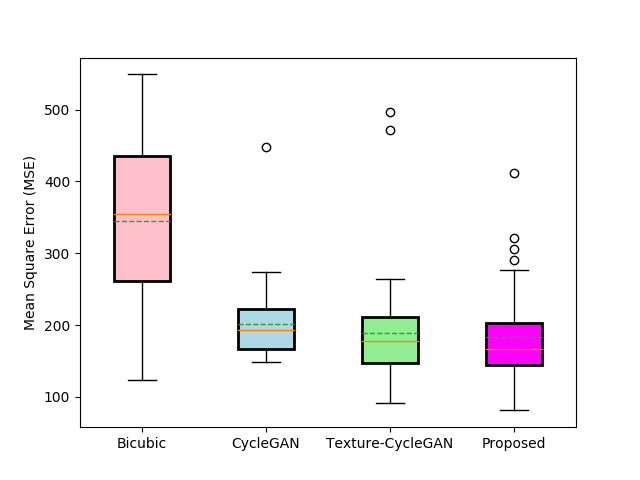}  \vspace{-2mm} \\
		(g) & 	(h)  & (i)  \\
		\end{tabular}
	\end{center}
% 	\vspace{-3mm}
\caption{Boxplot showing the three evaluation metrics of the different methods for the test patients. The red line represents the mean value, the green dashed line represents the median value and the circles represent outliers. (a), (b) and (c) are the task of MRI TI-Flair modality translation. (d), (e) and (f) are the task of MRI Flair-T2 modality translation. (g), (h) and (i) are the task of cross-protocol super resolution of diffusion MRI.}
\label{fig:fig3}
	\vspace{-4mm}
	\label{fig:fig3}
\end{figure*}

\subsection{Evaluation Metrics and Protocol}
Three metrics are used to evaluate the quantitative results of synthesis images in the aspects of similarity, human perception, and pixel-level differences. Given a target modality $G$ and a translated output $T$ generated by our algorithm. Three evaluation metrics are explained as follows:

{\flushleft\textbf{Structural similarity index measure (SSIM)}} is a method to predict the perceived quality of images on three comparison measurements between the luminance, contrast, and structure. In particular, it considers the strong inter-dependencies between pixels when they are spatially similar. It is well noted that in the visual scene, these dependencies hold considerable details about the structure of the objects.
% It considers the strong inter-dependencies between pixels especially when they are spatially close. It is well noting that these dependencies carry important information about the structure of the objects in the visual scene.

\begin{equation}
\rm{SSIM}(x,y)= \frac{(2\mu_{x}\mu_{y}+c_{1})(2\sigma_{xy}+c_{2})}{(\mu^{2}_{x}+\mu^{2}_{y}+c_{1})(\sigma^{2}_{x}+\sigma^{2}_{y}+c_{2})}
\end{equation}
where x and y are the windows of the translated map $T$ and the target modality $G$. $\mu_{x}$ and $\mu_{y}$ are the average of $x$ and $y$ while the $\sigma^{2}_{x}$ and $\sigma^{2}_{y}$ are the variance of $x$ and $y$. $\sigma_{xy}$ is the covariance of $x$ and $y$.

{\flushleft\textbf{Mean squared error (MSE)}} is an estimator calculates the mean of the squares of the errors between the target modality $G$ and the translated map $T$. 
\begin{equation}
\rm{MSE}=\frac{1}{mn}\sum_{0}^{m-1}\sum_{0}^{n-1}\left \| G(i,j)-T(i,j)      \right\|
\end{equation}
Where $n$ and $m$ are the width and height size of the target image. $i$ and $j$ are the pixel indexes of the translated image.
{\flushleft\textbf{Peak signal-to-noise ratio (PSNR)}} means the ratio between the maximum value of a signal and the power of distorting noise.
\begin{equation}
\rm{PSNR}=20\log_{10}(\frac{Max_{f}}{\sqrt{MSE}})
\end{equation}
Where $Max_{f}$ is the maximum signal value of the target image.

\begin{table*}[t]
% \vspace{-0.1cm}
\caption{Quantitative results of cross-protocol super-resolution, T1-Flair, and Flair-T2 modality translation tasks. The best results are highlighted in bold.}\label{tab:tab1}
\footnotesize
%\small  \footnotesize  \normalsize
\centering
\begin{tabular}{p{1.5cm}p{0.6cm}p{0.6cm}p{0.6cm}p{0.6cm}p{0.6cm}p{0.6cm}p{0.6cm}p{0.6cm}p{0.6cm}}
%\begin{tabular}{ccccc}
\toprule[1.5pt]
\multirow{1}{0pt}{Methods} & \multicolumn{3}{c}{Super-Resolution} & \multicolumn{3}{c}{T1-Flair translation} & \multicolumn{3}{c}{Flair-T2 translation}\\ \cline{2-10} 
 & PSNR & SSIM &MSE & PSNR & SSIM  & MSE & PSNR & SSIM  & MSE\\ \toprule[1.5pt]
Bicubic & 22.98  & 0.7315 & 344.9    & - & - & - & - & - & -\\ \hline
U-Net & -  & - & -    & 17.16 & 0.5012 & 1251.4 & 23.75 & 0.8231 & 274.2 \\ \hline
CycleGAN  & 25.18  & 0.7636 & 202.1     & 22.08 & 0.5727 & 438.9 & 21.69 & 0.8292 & 504.2\\ \hline
Texture-CycleGAN  & 25.65 & 0.7720 &  189.6 & 21.45 & 0.6786 &539.7 & 20.90 & 0.8079 &642.3 \\ \hline
Ours  & \textbf{25.75} & \textbf{0.7959} &\textbf{182.9} & \textbf{23.07} & \textbf{0.7238}  & \textbf{357.4} & \textbf{25.12} & \textbf{0.8855} & \textbf{220.5}  \\ \toprule[1.5pt]
\end{tabular}
% \vspace{-7mm}
\end{table*}

\section{Experiments and Analysis}
To assess the performance of our multi-scale texture learning algorithm in tackling the inter-scanner and inter-protocol discrepancy on Magnetic Resonance Imaging (MRI), we extensively design three different tasks: 1) cross-protocol super resolution of diffusion MRI. 2) MRI TI-Flair modality translation. 3) MRI Flair-T2 modality translation. These applications are very vital and promising in MRI harmonization from different scanners, and in the data filling and modality synthesis when it fails to collect fully complete multi-modality MR images.

\subsection{Implementation details}
For all the experiments, we set $\lambda_{cyc}=10 $ in Eq. 9. We train our model 100 epochs with ADAM optimize from scratch and the batch size is set as 1. The learning rate is 0.0002 initially and is divided by 2 every 50 epochs. We implement experiments with PyTorch using an NVIDIA TITAN X GPU.

%%%--------------
\begin{figure*}[t]
\centering
\includegraphics[width=0.85\textwidth]{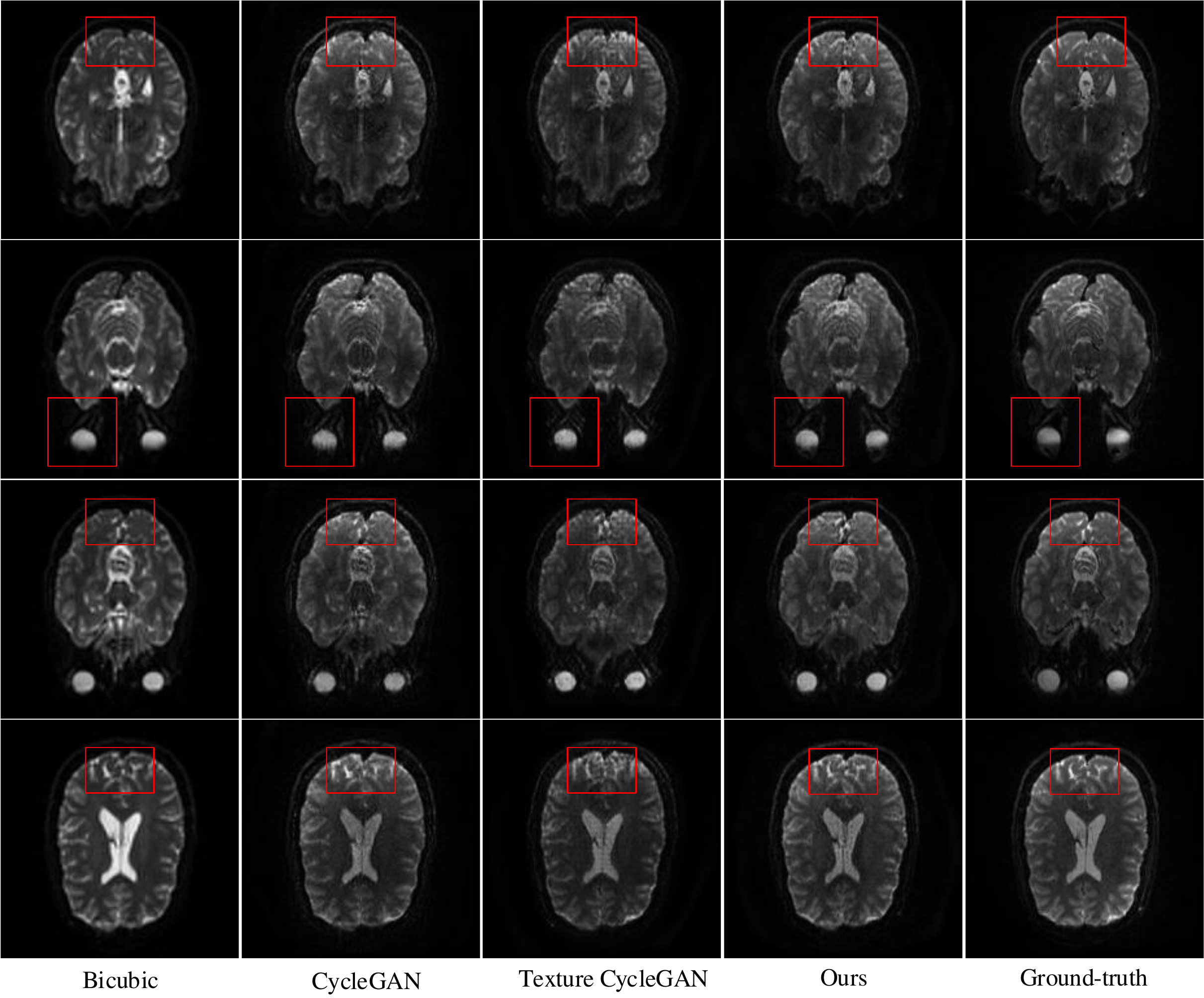}
\caption{Qualitatively comparison with state-of-the-art methods of cross-protocol super-resolution on the diffusion MRI data (MUSHAC). Best viewed by zooming in on the screen. }\label{fig:fig4}
\end{figure*}
%%%------------------

\subsection{Comparisons with State-of-the-Art Methods}
In order to evaluate the performances of our algorithms, we compare them with the state-of-the-art methods qualitatively and quantitatively. More specifically, the traditional bicubic interpolation is the basic baseline for super-resolution task and U-Net is chosen as the pixel-level translation baseline for the image translation task. The CycleGAN \cite{gan10} and the texture-CycleGAN combined with the single-scale texture transfer based on a pre-trained VGG-18 net \cite{gatys2016image} are chosen as the performance baselines. To ensure the fair comparisons with STOA methods, we train these methods on the same dataset by the open-source implementations as well as the recommended hyper-parameters in the original publication.

% To ensure faithful representation of the baseline methods, 

\subsection{Quantitative Analysis}
The quantitative evaluations of the network using the structural similarity (SSIM), the peak signal-to-noise ratio (PSNR), and mean squared error (MSE) scores for the cross-protocol super resolution of diffusion MRI, T1-Flair, and Flair-T2 modality translation are listed in Table \ref{tab:tab1}.

{\flushleft \textbf{Cross-protocol super resolution of diffusion MRI:}} As shown in Table \ref{tab:tab1}, our multi-scale texture GAN achieves better results than the single-scale texture-CycleGAN and 2.77dB higher than the traditional bicubic method. It's worth noting that our network has no difference with the Texture-CycleGAN network except for the difference in the texture transfer method. These results demonstrate that the proposed multi-scale texture mechanism makes a significant contribution to performance improvement compared to the basic CycleGAN without any texture transfer operation. In comparison with single-scale texture transfer, our multi-scale texture algorithm performs better in the task of cross-protocol super resolution of diffusion MRI. 
Figure \ref{fig:fig3}(g-i) show the distributions and outliers of the three metrics obtained by four schemes for the test dataset. We can see that our multi-texture GAN produces fewer outliers and therefore, provides more stable and more accurate quantitative results, compared to the other three methods. Figure \ref{fig:fig6}(c) provides that the histogram and its corresponding correlation metric result compared with the target image. It demonstrates that the pixel-level intensity distribution calculated by  our model is closer to the image of higher-resolution scanner than other models. 

%%---------
\begin{figure*}[t]
\centering
\includegraphics[width=0.9\textwidth]{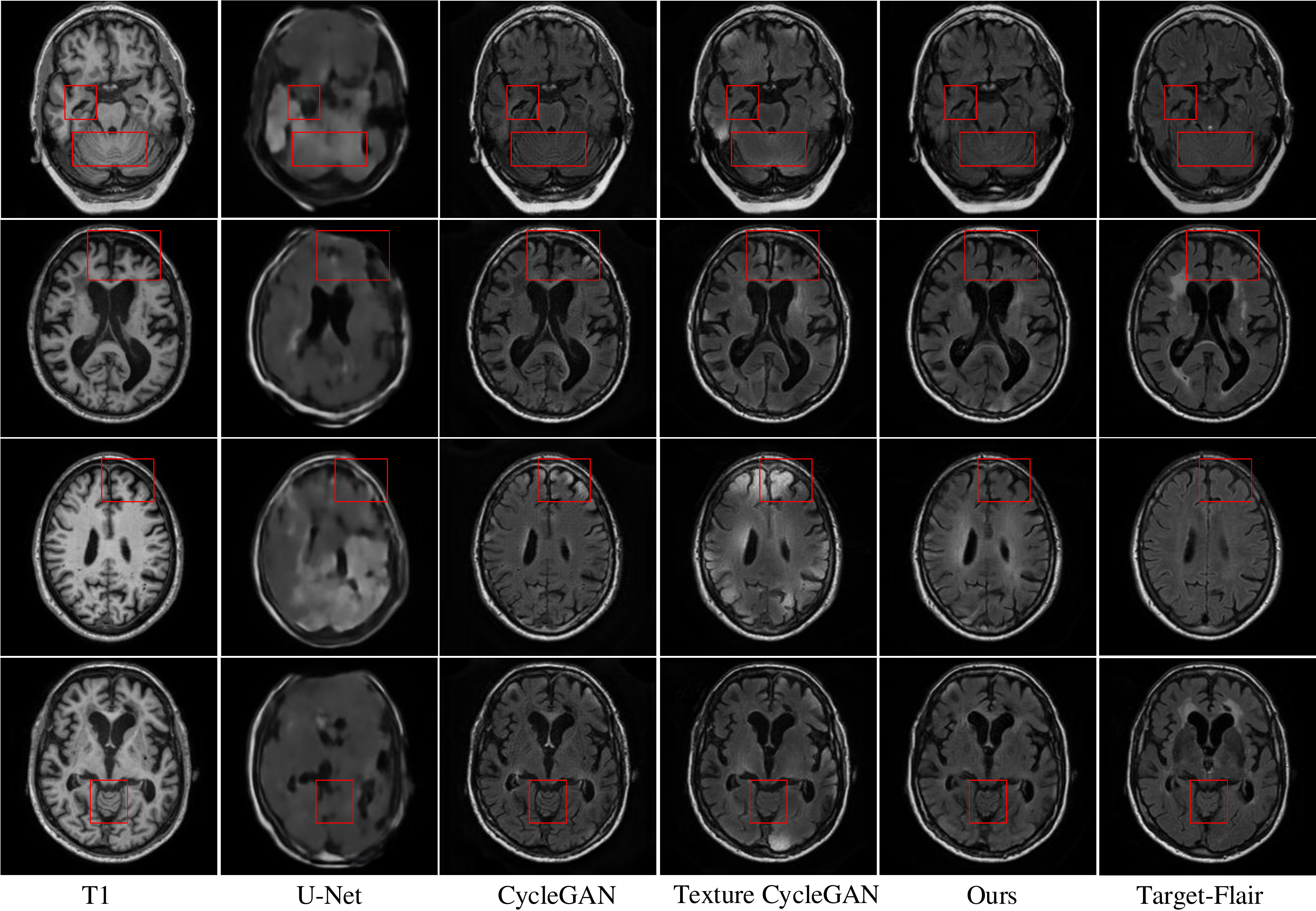}
\caption{Qualitatively comparison with state-of-the-art methods of T1-Flair modality translation. Best viewed by zooming in on the screen.}\label{fig:fig5}
% \vspace{-0.7cm}
\end{figure*}
%%------

{\flushleft \textbf{T1-Flair modality translation:}}
To verify the effectiveness of the proposed network in the task of modality translation, the quantitative results on TI-Flair modality translation are reported in Table \ref{tab:tab1}. Our multi-scale texture GAN achieves significantly better results in PSNR and SSIM (0.99 dB and 0.1511 higher than the CycleGAN). It is noticed that the single texture transfer deteriorates the result in PSNR while the multi-scale texture transfer mechanism greatly improves the performance both in PSNR and SSIM. It indicates that a multi-scale texture mechanism is more robust and effective in the task of the modality translation than the single-scale texture. The U-Net method, which focuses on the pixel-level information transformation between the T1 and Flair modality, performs extremely poorly (17.16 dB) in the task of the T1-Flair translation. The supporting reason is that the T1 and Flair modalities have weak pixel-similarity since TI modality contains more anatomy structure and Flair modality consists of more lesion knowledge. Figure \ref{fig:fig3}(a-c) shows the distributions of the performances in terms of PSNR, SSIM, and MSE on the test patients using four different methods. It illustrates that our algorithm can provide more stable and higher accurate results. The pixel intensity distribution of our algorithm is shown in Figure \ref{fig:fig6}(a). Compared with others, our model has a higher histogram correlation with the target modality.

\begin{figure*}[t]\scriptsize
	\begin{center}
		\tabcolsep 1pt
		\begin{tabular}{@{}ccc@{}}
			\includegraphics[width=0.33\textwidth]{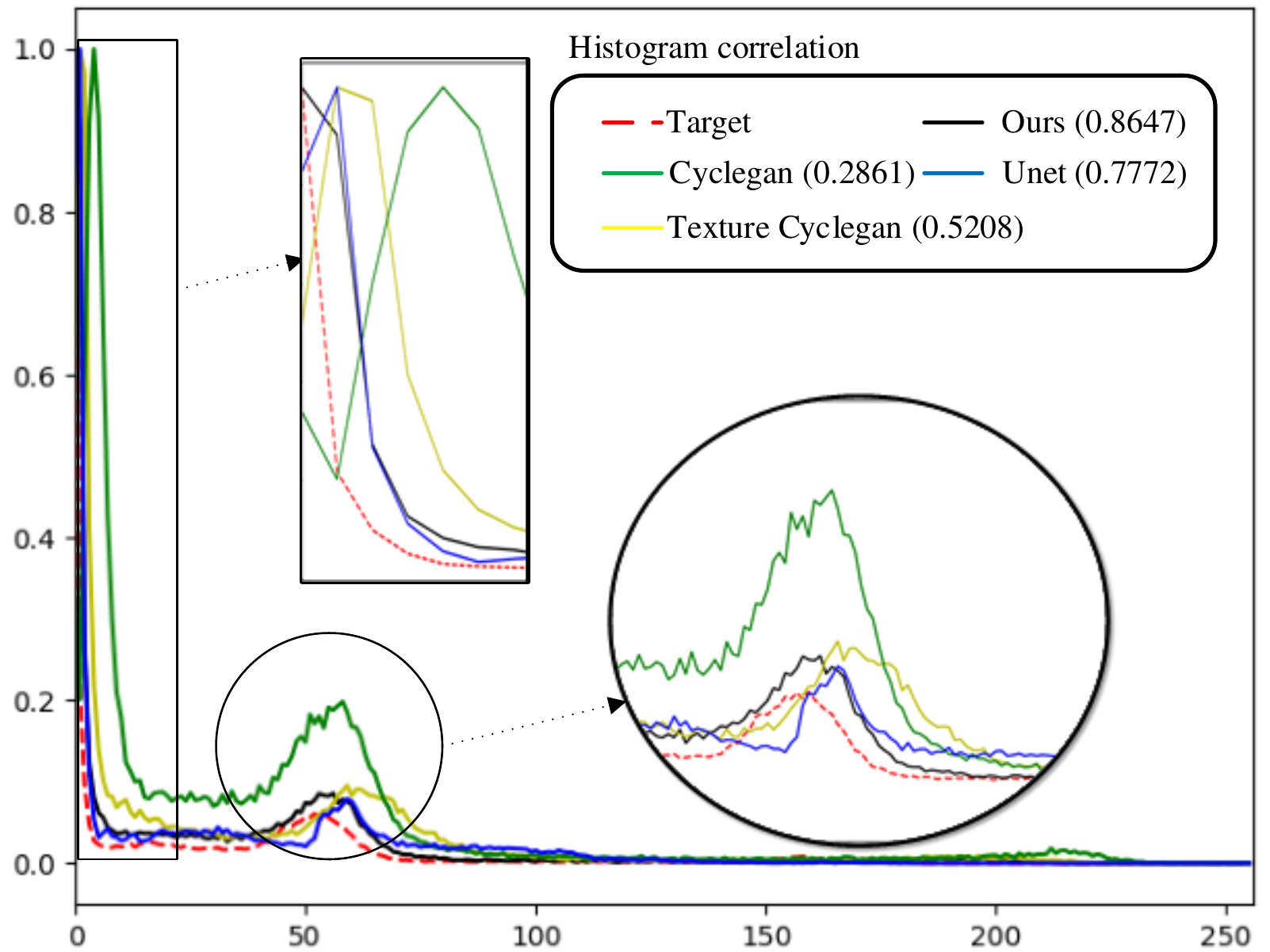}  & 
			\includegraphics[width=0.33\textwidth]{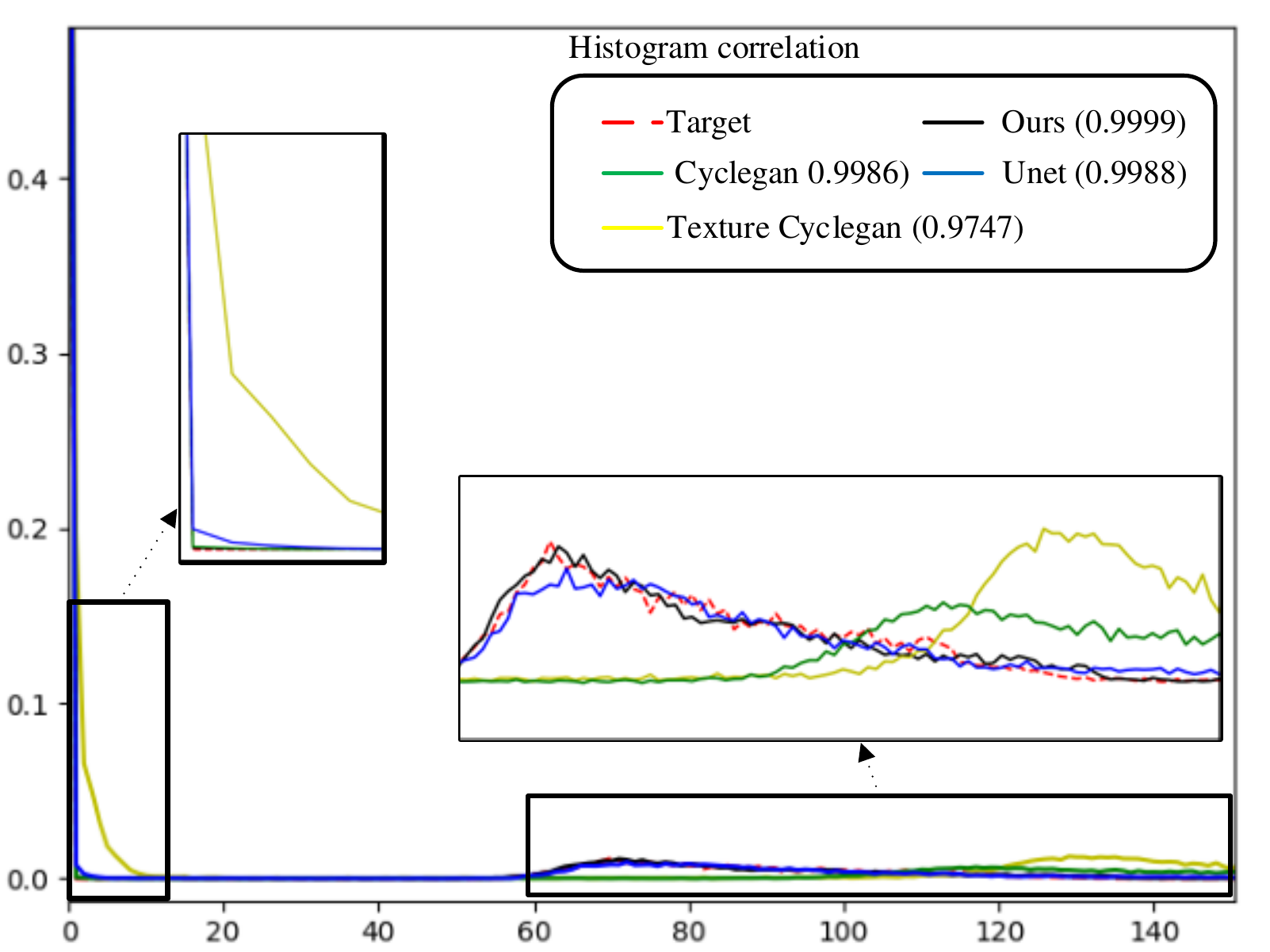} & 
			\includegraphics[width=0.33\textwidth]{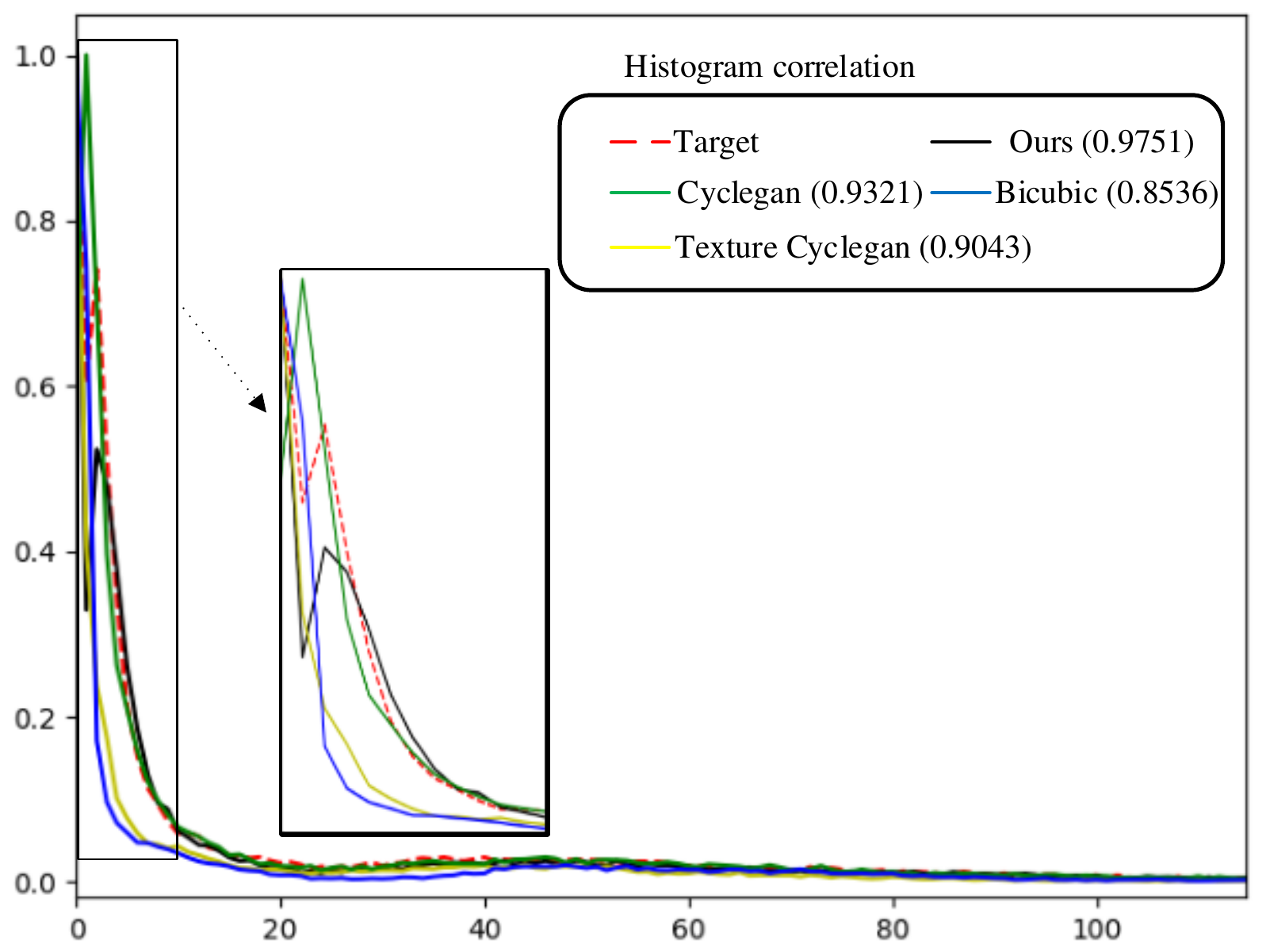} \\
			 (a)   & (b) & (c) \vspace{1pt} \\
		\end{tabular}
	\end{center}
% 	\vspace{-3mm}
\caption{Histograms showing the pixel intensities distribution of different methods for the test patients compared with the distribution of the target modality. The histogram correlation, a metric to evaluate the distribution similarity, is shown here to demonstrate the superiority of our algorithm. (a) is the task of MRI T1-Flair modality translation and (b) is the task of MRI Flair-T2 modality translation. (c) is the task of cross-protocol super resolution of diffusion MRI.}
% 	\vspace{-4mm}
	\label{fig:fig6}
\end{figure*}

\begin{figure*}[htbp]
\centering
\includegraphics[width=0.9\textwidth]{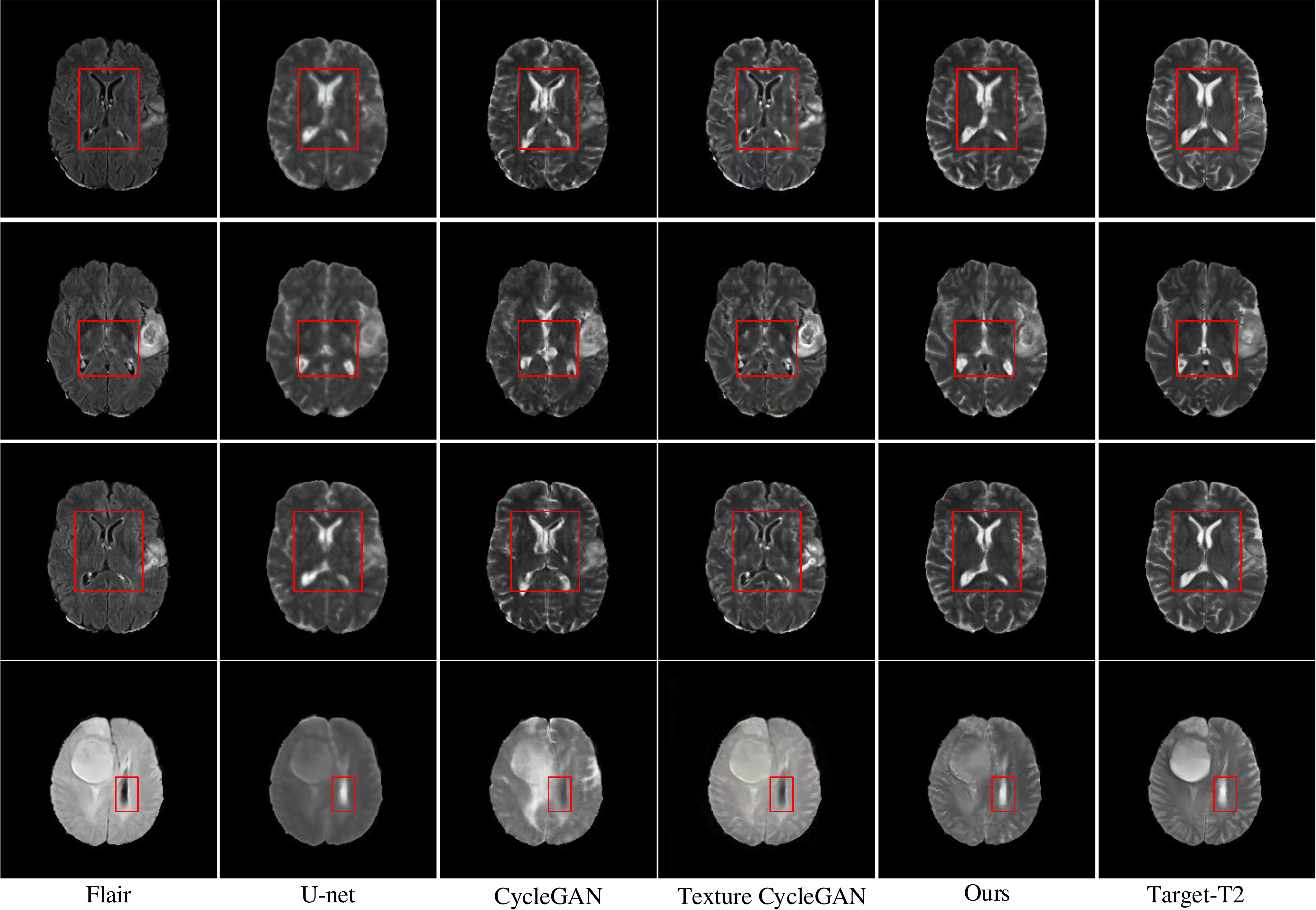}
\caption{Qualitatively comparison with state-of-the-art methods of Flair-T2 modality translation. Best viewed by zooming in on the screen. }\label{fig:fig7}
% \vspace{-0.7cm}
\end{figure*}

{\flushleft \textbf{Flair-T2 modality translation:}}
For the task of Flair-T2 modality translation, the quantitative details are also listed in Table \ref{tab:tab1}. It indicates that our multi-scale texture algorithm (25.12dB) significantly outperforms the  U-Net (23.75dB), original CycleGAN (21.69dB) and single-scale texture CycleGAN (20.90dB). However, the pixel-level U-Net is the second-best method (PSNR) in this task while its quantitative performance in the task of T1-Flair modality translation is the worst. We can note that the pixel-level U-Net performs well when the Flair and T2 modalities have a high pixel-level similarity. From the results' comparisons, our Multi-Texture GAN is still the best model in the high pixel-level similarity transformation when other texture-based methods (CycleGAN and single-texture CycleGAN) fail to provide satisfactory results. Figure \ref{fig:fig3}(d-e) shows the distributions of three metrics on the test dataset by four models and Figure \ref{fig:fig6}(b) shows the pixel intensity distribution on the task of Flair-T2 modality translation.

\subsection{Qualitative Evaluation}
{\flushleft \textbf{Cross-protocol super resolution of diffusion MRI:}}
For the qualitative analysis, the cross-protocol translation from the standard Prisma scanner to the state-of-the-art Connectom scanner is shown in Figure \ref{fig:fig4}. The images of STOA Connectom scanner have higher-resolution with more details but do not show the high contrast foreground. We also find that our algorithm restores higher-resolution details and low-contrast foreground similar to the ground-truth compared with other models.
As shown in the red box, our method rehabilitates structure information much closer to the Ground-truth compared with others.

{\flushleft \textbf{T1-Flair modality translation:}}
The visual quality evaluation of T1-Flair modality translation is shown in Figure \ref{fig:fig5}.  
For the low pixel-level similarity translation, the U-Net generates blurred images and fails to recover clear structure information. Compared with single-scale texture, the multi-scale texture mechanism performs much better in contrast and texture transformation between T1 and Flair modality. The red box shows that our method rehabilitates better texture (e.g., contrast) and structural (e.g., edges, anatomy) information. 

{\flushleft \textbf{Flair-T2 modality translation:}}
The qualitative results of our methods in Flair-T2 modality translation in Figure \ref{fig:fig7}. It is worth noting that U-Net gets a good quantitative result but fails to rehabilitate high-resolution images compared with other methods. It restores a high-pixel similarity image as same as the target modality but the high-resolution details are ignored. For lesion-based translation, our method not only inherits the lesion information from source modality (Flair) but also optimizes the lesion information (including the contrast, texture, and shape of the lesion) according to target modality (T2). The red box shows our multi-texture GAN can recover better texture (e.g., contrast and edges), structural and lesion information. 

% We find that our algorithm restores high-resolution details similar with the ground-truth. For the modality translation, it can be observed that our proposed method transfers more texture, and clearer edges and contrast close to the T1 modality. 

% one-senstence lesion information 

\section{Discussion}
This section aims to analyze the role of our multi-scale texture mechanism,  the main difficulty of modality translation (texture and lesion-related knowledge), and its potential solutions.
\subsection{\textbf{Multi-texture GAN:} With/without Multi-scale texture}
Our multi-scale texture mechanism is a plug-and-play module that can be embedded into any existing framework. Specifically, a multi-scale matching conducted in the neural space allows the model to benefit more from the semantically related reference patches.
After adding the multi-texture mechanism, it shows the superiority in restoring a more high-resolution structure (i.e., edges and anatomy), texture (i.e., contrast and pixel intensities), and lesion information. The multi-scale texture mechanism bridges the gap between the source and target translation by shortening the distance of the texture discrepancy and then make it easier for the inter-scanner and inter-protocol translation.

% Semantic-related translation
% Pixel intensity distribution of synthesis image

\subsection{\textbf{Multi-texture GAN:} Multi-scale texture vs single-scale texture}
Our multi-scale texture translation enriches synthesis images by adaptively transferring the high-level (e.g., semantic or lesion-related) and low-level (e.g., contrast) texture from reference modality according to the textural similarity. To dismiss the constraint on the global structure of the reference images, we match the input patches with the down-up scale of reference and then calculate the texture similarity to select the highest similarity patch for the feature swapping. Instead of it, the single-scale texture just separates and recombines the image content and style without the feature matching and similarity calculation. Consequently, it may generate some misleading texture information including the contrast and lesion-based information on the final synthesis images. This conclusion supports the result that the CycleGAN without any texture translation performs better than the single-scale texture in the tasks of F1-Flair and Flair-T2 translation. Overall, our multi-scale texture algorithm is a robust and accurate solution to rehabilitate synthesis images with finer texture and structural information while the single-scale texture deteriorates the results in some cases.  

\subsection{\textbf{Multi-texture GAN:} Pixel-level vs texture-level translation}
It is well-known that the T1 modality has strong signals in the area of anatomy, protein-rich fluid, and paramagnetic or diamagnetic substance. Conversely, T2 modality contains strong signals on the area of pathology and water content as in edema and tumor, and similarly, T2-Flair modality has strong signals on the area of the lesion. 
The information discrepancy between modalities determines the translation difficulty. Due to the large discrepancy between T1-Flair modalities, the pixel-level U-Net performs much worse than the GAN-based texture networks. On the other hand, the pixel-level U-Net obtains a satisfactory result for the high pixel-similarity modality translation (e.g., Flair-T2). 
Conversely, the GAN-based texture method can get relatively better results when the texture and content discrepancy of modalities translation is large. Finally, it is worth noting that our multi-texture performs much better than other pixel-level or texture-level translation methods in both the high or low pixel-similarity translation tasks.

\section{Conclusion}
Multi-scale texture GAN is a new framework for image translation by incorporating multi-scale adaptive texture priors obtained by the neural space. It builds upon the widely utilized CycleGAN framework and embeds the texture priors extracted by a multi-scale texture mechanism to bridge the gap between the source and target domain. Specifically, we match texture features in a multi-scale scheme, calculate the similarity score, and then pick out the most semantic-related features for the feature swapping. The comprehensive experimental results of cross-protocol super-resolution and cross-modality translation have shown that our method achieves superior performance compared with the SOTA algorithms.

\ifCLASSOPTIONcaptionsoff
  \newpage
\fi



%\bibliographystyle{IEEEtran}

\bibliographystyle{IEEEtran}
\bibliography{rnndeblur}

\end{document}